\documentclass[twocolumn,aps,pre,amsmath,amssymb,superscriptaddress,nofootinbib]{revtex4}
\usepackage {CJK}
\usepackage{graphicx}
\usepackage{bbold}
\usepackage{float}
\usepackage{color}
\usepackage{relsize}

\def\be{\begin{eqnarray}}
\def\ee{\end{eqnarray}}
\def\ben{\begin{eqnarray*}}
\def\een{\end{eqnarray*}}
\def\bes{\begin{subequations}}
\def\ees{\end{subequations}}

\def\nn{\nonumber}

\def\la{\langle}
\def\ra{\rangle}

\def\Tr{{\rm Tr}}
\def\l{\left}
\def\r{\right}
\newcommand{\wig}[1]{\mathrel{\hbox{\hbox to 0pt{\lower.6ex\hbox{$\sim$}\hss    }\raise.4ex\hbox{$#1$}}}}

\begin{document}
\begin {CJK*} {GB}{}

\title{Non-Adiabatic Quantum Molecular Dynamics with Detailed Balance}
\author{J\'er\^ome \surname{Daligault}}
\email{daligaul@lanl.gov}
\author{Dmitry \surname{Mozyrsky}}
\affiliation{Theoretical Division, Los Alamos National Laboratory, Los Alamos, NM 87545, USA}

\begin{abstract}
We present an approach for carrying out non-adiabatic molecular dynamics simulations of systems in which non-adiabatic transitions arise from the coupling between the classical atomic motions and a quasi-continuum of electronic quantum states.
Such conditions occur in many research areas, including chemistry at metal surfaces, radiation damage of materials, and warm dense matter physics.
The classical atomic motions are governed by stochastic Langevin-like equations, while the quantum electron dynamics is described by a master equation for the populations of the electronic states.
These working equations are obtained from a first-principle derivation.
Remarkably, unlike the widely used Ehrenfest and surface-hopping methods, the approach naturally satisfies the principle of detailed balance at equilibrium and, therefore, can describe the evolution to thermal equilibrium from an arbitrary initial state.
A practical algorithm is cast in the form of the widely used fewest switches surface-hopping algorithm but with switching probabilities that are not specified {\it ad-hoc} like in the standard algorithm but are instead derived.
\end{abstract}


\date{\today}

\maketitle
\end{CJK*}

\section{Introduction} \label{Section_1}

Mixed quantum-classical dynamics methods are extensively used for simulating systems in many areas of the physical and chemical sciences \cite{Tully1998}.
These methods give a classical treatment to atomic motions, while retaining a detailed quantum mechanical description of electrons.
This considerably reduces the formidable computational cost entailed by a complete quantum mechanical description.
The development of mixed quantum-classical dynamics methods beyond the Born-Oppenheimer approximation, by which electrons follow adiabatically the classical atomic motions, has been a topic of continuous research interest for several decades \cite{Tully2012}.
In general, atomic motions can induce transitions between electronic states, which, in turn, can alter the forces acting on the classical particles.
Such non-adiabatic effects are ubiquitous and diverse \cite{Tully2012}, and the self-consistent incorporation of feedback between the quantum and classical degrees of freedom is highly non-trivial \cite{Tully1990,Stella2007}.
A well-known difficulty is to ensure that the principle of detailed balance, according to which transitions between any two states take place with equal frequency in either direction at equilibrium, be satisfied \cite{Tully2012,Parandekar2005}.
Failure to satisfy detailed balance introduces a bias and systematically skews the dynamics away from thermal equilibrium.

In this paper, we present an approach for carrying out non-adiabatic quantum-classical molecular dynamics simulations of systems in which non-adiabatic transitions arise from interactions between the motion of the classical degrees of freedom and a quasi-continuum of quantum states.
Unlike the popular Ehrenfest method and Tully's trajectory surface-hopping method \cite{Parandekar2005,Bastida2006}, the new scheme naturally satisfies detailed balance.
The atomic motions are governed by stochastic Langevin-like equations, while the electron dynamics is described by a master equation for the populations of the electronic states.
The scheme can thus properly describe the irreversible evolution of an isolated system from an arbitrary initial state to a state of thermal equilibrium.
At equilibrium, the transition rates between electronic states satisfy the detailed balance relations, while the non-adiabatic  forces acting on the ions satisfy the fluctuation-dissipation relation.

There exists a large number of systems in which non-adiabatic effects can arise as a consequence of the coupling between the atomic motions and a quasi-continuum of electronic states \cite{Tully2012}.
The situation, which differs from the more commonly discussed case of a handful of strongly coupled energy levels, is in clear conflict with the Born-Oppenheimer criterion that the states be widely separated in energy.
Here excitations of arbitrarily low energy are available to couple with the classical motions.
Such couplings are known to significantly affect dynamical processes such as adsorption, dissociation, and catalytic reaction at metal surfaces \cite{Wodtke2004,Askerka2016}.
In solids, atomic diffusion of impurities in metals \cite{Li1995} and the radiation damage processes induced by energetic particles \cite{Race2010,Correa2012} involve important nonadiabatic couplings with host electrons.
In warm dense matter \cite{IPAM2012}, not only there is a quasi-continuous density of electronic states at the Fermi level, but the volume of available, unoccupied states can be large since electrons are partially degenerate.
Non-adiabatic couplings could potentially affect dynamical ionic properties even at equilibrium; without doubt, non-adiabatic couplings must be accounted for to calculate quantities of current experimental interest, such as temperature equilibration rates \cite{Pradhan2016,Cho2016}.
The list is not limited to bulk systems, as finite systems can also display a dense manifold of electronic states \cite{Valero2012} (see \cite{Tully2012} for an extensive list).

Our scheme results from a fairly long mathematical derivation.
To ease the presentation, in Sec.~\ref{Section_2}, we first outline the scheme, enumerate its salient properties and propose an algorithm.
We then proceed in Sec.~\ref{Section_3} with the complete derivation.
For the readers who plan to skip the proofs given in Sec.~\ref{Section_3}, we remark that, unlike other derivations of mixed quantum-classical schemes such as Ehrenfest, ours does not treat from the outset the atomic positions and momenta as classical parameters in the equations of electrons; this is indeed known to be at the origin of the breakdown of detailed balance \cite{BlumBook,Abragam1961}.
Instead, it ensures that the canonical commutation relations of atomic variables in these equations are satisfied.

\section{Outline of the scheme} \label{Section_2}

\subsection{The scheme}

{\it Definitions and assumptions.} Following standard notations, let $r$ designate the three-dimensional cartesian positions of electrons (mass $m$) and $R$ the atomic positions (mass $M$) \cite{Note_extension}.
Below, $N$ denotes the total number of atoms and $R=\{R_\alpha\}_{\alpha=1,\dots,3N}$ denotes the set of all atomic positions.
The total Hamiltonian describing the system is
\be
\hat{H}(r,R)=-\frac{\hbar^2}{2M}\nabla_R^2+\hat{H}_e(r,R) \label{complete_Hamiltonian}
\ee
where $\hat{H}_e(r,R)$ is the electronic Hamiltonian for fixed atomic position and $\nabla_R=\partial/\partial R$.
For simplicity of exposition, we assume that there is no external time-dependent potential acting on the system; we consider situations in which the system is either in thermal equilibrium, or initially excited and then let to freely evolve and relax.
We also assume that $m\ll\! M$ and that the atomic velocities are large enough that the atomic de Broglie wavelengths are smaller than the characteristic variation length scales of interactions; thus the atomic motions can be described by classical-like trajectories.
If at time $t$ the atomic positions are $R(t)$, we define the basis of adiabatic wave functions $|i(R(t))\rangle$ as the eigenfunctions of $\hat{H}_e\l(r,R(t)\r)$ \cite{Tully1998}, i.e.
\be
\hat{H}_e\l(r,R(t)\r)|i(R(t))\rangle&=&\epsilon_i(R(t))|i(R(t))\rangle\,.\label{adiabatic_basis}
\ee
From now on, we often omit writing explicitly the dependence on $R(t)$ of the adiabatic basis and related quantities in order to avoid cluttering the mathematical expressions.
We define the non-adiabatic couplings $d_{ij}=\left\langle i\left| \nabla_R\right| j\right\rangle=-d_{ji}^*$ and $f_{ij}=\left\langle i\left| -\nabla_R\hat{H}_e(r,R(t))\right| j\right\rangle=\epsilon_{ij}d_{ij}$ with $\epsilon_{ij}=\epsilon_i-\epsilon_j$ \cite{Tully1998}.

As discussed above, we consider physical systems in which the electronic energy states $\epsilon_i(R)$ form a continuum or a manifold of infinitesimally separated electronic excitations. A large number of electronic states implies the existence of a short time scale $\tau_c$ (discussed below) arising from the rapid fluctuations of coherences $\rho_{ij}(t)=\Tr\left[\hat{\rho}(t)|i\ra\la j|\right]$ with $i\neq j$ (with $\hat{\rho}$ the total density operator of the system), which, in turn, affect the atomic motions in the form of a rapidly fluctuating force. In our scheme, coherences are not treated explicitly; their influence is treated statistically and is responsible for the stochastic nature of the classical atomic motions discussed below.
Instead, our approach describes the evolution of the atomic positions $R(t)$ and of the electronic populations $P_i(t)=\Tr\left[\hat{\rho}(t)|i\ra\la i|\right]$ on a time scale coarse-grained over $\tau_c$.

{\it Working equations of the scheme.} Each atomic position $R_\alpha$  satisfies the stochastic equation
\be
M\ddot{R}_{\alpha}(t)=F_{\alpha}^{BO}(t)-M\sum_{\beta=1}^{3N}{\gamma_{\alpha,\beta}\dot{R}_{\beta}(t)}+\xi_{\alpha}(t)\,\label{Langevinequation}
\ee
with initial conditions $R_\alpha(0)$ and $\dot{R}_\alpha(0)$ at initial time $t=0$.
In these equations, dropping the time variable,
\be
F_\alpha^{BO}=-\sum_{i}{P_if_{ii}^\alpha}\,,\label{F_BO}
\ee
is the adiabatic Born-Oppenheimer force defined with respect to the adiabatic states (\ref{adiabatic_basis}) at time $t$, i.e. the average over all states $i$ of the forces $-f_{ii}^\alpha$ weighted by the occupation number $P_i$.
The remaining two terms result from the non-adiabatic couplings:
a sum of friction forces $-M\sum_{\beta=1}^{3N}{\gamma_{\alpha,\beta}\dot{R}_{\beta}(t)}$ with friction coefficients
\be
\gamma_{\alpha,\beta}=-\frac{\pi}{M}\sum_{i\neq j}{\frac{P_i-P_j}{\epsilon_{ij}}f_{ij}^{\alpha}f_{ji}^{\beta}\,{\cal{L}}(\epsilon_{ij},\Gamma_{ij})}\,, \hspace{0.25cm}\label{friction_gamma}
\ee
which describes the systematic effect of non-adiabatic transitions on the atomic motions and damps the velocities over a characteristic time $T_\gamma=1/\gamma$;
and a delta-correlated Gaussian random force $\xi_\alpha(t)$ satisfying \cite{note_on_delta_time}
\be
\ll\!\xi_{\alpha}(t)\!\gg=0\quad,\quad\ll\!\xi_{\alpha}(t) \xi_{\beta}(t')\!\gg=B_{\alpha,\beta}\delta(t-t')\,, \label{xi_alpha}
\ee
with
\be
B_{\alpha,\beta}=\pi\sum_{i\neq j}{(P_i+P_j)f_{ij}^{\alpha}f_{ji}^{\beta}{\cal{L}}(\epsilon_{ij},\Gamma_{ij})}\,, \label{fluctuation_dissipation}
\ee
which describes the fluctuations of non-adiabatic forces around their average values, and varies over a short time scale of the order of $\tau_c$ (see below).
The Lorentzian
\be
{\cal{L}}(\epsilon_{ij},\Gamma_{ij})=\frac{1}{\pi}\frac{\hbar^2\Gamma_{ij}}{(\epsilon_{ij})^2+(\hbar\Gamma_{ij})^2}\label{call}
\ee
describes the energy conservation, corrected by the broadening of the transition due to the finite lifetime of the coherence between states $i$ and $j$ (recall that ${\cal{L}}(\epsilon,\Gamma)\sim \delta(\epsilon/\hbar)$ as $\Gamma\to 0$).
The inverse lifetime $\Gamma_{ij}$ is found  from the self-consistency equation,
\be
\Gamma_{in}&=& {2\hbar^2\over (M d_{in}\cdot V)^2}\sum_{j\neq j'}
\frac{(d_{in}\cdot f_{j'j})^2\Gamma_{jj'}\,P_j}{(\epsilon_{jj'})^2+(\hbar\Gamma_{jj'})^2}\,.\label{gammaself1}
\ee
where the dot sign denotes the scalar product and $V(t)=\dot{R}(t)$ the atomic velocities.

We thus recover that the effect of non-adiabatic couplings is analogous to that of collisions undergone by a heavy Brownian particle immersed in a fluid of light particles.
In the latter, the Brownian motion appears erratic over a time scale $\tau_c$ of several successive collisions, while a much longer time scale $T_\gamma=1/\gamma$, or equivalently a significant number of collisions, is required to move appreciably the Brownian particle from its inertial motion.
In our case, $\tau_c$ and $T_\gamma$ can be identified as follows.
For simplicity of notations, consider the case of a system at thermal equilibrium at temperature $T$ and one atomic degree of freedom.
The friction coefficients (\ref{friction_gamma}) are then given by the Green-Kubo formula (\ref{friction_eq}) below in terms of the time correlation functions of the adiabatic forces.
In the presence of a quasi continuum of state, it is easily seen that these correlations decay very rapidly with time: $\langle \delta \hat{F}(t) \delta \hat{F}(0)\rangle\approx \la \delta\hat{F}^2\ra e^{-t/\tau_c}$ where $\tau_c$ is the correlation time of the nonadiabatic force; this implies $T_\gamma=1/\gamma=M^2v_{th}^2/\la \delta\hat{F}^2\ra \tau_{c}$ with the thermal velocity $v_{th}=\sqrt{k_BT/M}$.
The time $T_\gamma\!\gg\tau_c$ characterizes the time necessary for the cumulative effect of non-adiabatic electron-ion interactions to damp the atomic velocities.
The condition $\tau_c\ll\! T_\gamma$ on which our treatment relies, writes $\sqrt{\la \delta\hat{F}^2\ra }\,\tau_c/Mv_{th}\ll\! 1$.
It expresses that the evolution due to non-adiabatic couplings has a very weak effect during the correlation time $\tau_c$, in analogy with the weak effect of individual collisions on a classical Brownian particle.

In the past, several mathematical derivations of Eqs.(\ref{Langevinequation}-\ref{fluctuation_dissipation}) have been published at different levels of mathematical rigor \cite{dAgliano1975,HeadGordonTully1995,DaligaultMozyrsky2007,Lu2012,Dou2017}.
These works, however, treated the electronic subsystem as a reservoir, i.e. they assumed that the electronic subsystem is not modified by its coupling with atoms, and remains in a steady state.
The present work goes beyond this limitation and, as described below, gives an explicit treatment of the modifications of the state of electrons resulting from the non-adiabatic couplings between the electronic and atomic degrees of freedom.
In the equilibrium limit, the present results are in perfect agreement with previous works.

Our approach describes the electronic dynamics in terms of the evolution of populations of adiabatic states according to the master equation
\be
\frac{dP_i}{dt}&=&\sum_{a}{\left\{W_{ia}P_a-W_{ai}P_i\right\}}\,. \label{dPidt}
\ee
The first sum describes both the gain of state $i$ due to non-adiabatic transitions induced by the atomic motions from other states $a$, and the loss due to non-adiabatic transitions from $i$ into other states $a$.
The corresponding transition rates are
\be
W_{ia}=2\pi [d_{ia}\cdot V]^2\,e^{-{\epsilon_{ai}d_{ia}^2\over 2M[d_{ia}\cdot V]^2}}\,  {\cal{L}}\l(\epsilon_{ia},\Gamma_{ia}\r) \,.\label{w_ia}
\ee
The term $2\pi|d_{ia}\cdot V|^2{\cal{L}}$ is similar to the expression that one would obtained with a Fermi golden rule calculation by treating the atomic subsystem as an external disturbance on the electronic subsystem.
The exponental term $e^{-{\epsilon_{ai}d_{ia}^2\over 2M[d_{ia}\cdot V]^2}}$ results from the careful treatment of the quantum commutation relations of atomic variables in the equation of evolution of electronic populations; when the atomic positions are treated purely classically from the outset, as in the Ehrenfest method, this term equals unity.
As we shall discuss below, with this term, the rates (\ref{w_ia}) satisfy the principle of detailed balance.

\subsection{Salient properties}

We now discuss the key properties of the scheme.\\
{\it (i) Equilibrium limit, detailed balance and fluctuation-dissipation.}
Given a temperature $T$, the classical and quantum Boltzmann distributions $f_{eq}(R,V)=\exp\left[-\frac{1}{k_BT}\left(\frac{MV^2}{2}+\phi_{B0}(R)\right)\right]/{\cal{Z}}_{cl}$ and $P_i^{eq}(R)=e^{-\epsilon_i(R))/k_BT}/{\cal{Z}}_q$ with partition functions ${\cal{Z}}_{cl}=\iint{dRdV f_{eq}(R,V)}$ and ${\cal{Z}}_q=\sum_i{e^{-\epsilon_i(R)/k_BT}}$, and BO potential $\phi_{BO}=-k_BT\ln {\cal{Z}}_q$, constitute an equilibrium solution of the dynamics governed by Eqs.(\ref{Langevinequation})-(\ref{dPidt}).
Indeed, Eqs.(\ref{friction_gamma}) and (\ref{fluctuation_dissipation}) with $P_i=P_i^{eq}$ give the celebrated fluctuation-dissipation relation
\be
\ll\!\xi_\alpha(t) \xi_{\beta}(t')\!\gg&=&2Mk_BT\gamma_{\alpha,\beta}\delta(t-t')
\ee
with
\be
\gamma_{\alpha,\beta}=\frac{1}{2Mk_BT}{\rm Re}\int_{-\infty}^\infty{\langle \delta F_\alpha(t) \delta F_{\beta}(0)\rangle dt}\,. \label{friction_eq}
\ee
where $\delta{F}_\alpha=F_\alpha-F_\alpha^{0}$ with the electron-ion force $F_\alpha=\partial\hat{H}_e(r,R)/\partial R_\alpha$ and its diagonal part $F_\alpha^{0}$ in the adiabatic basis.
Equation (\ref{Langevinequation}) reduces to the traditional Langevin equation, which is known to yield the stationary distribution function $f_{eq}(R,V)$ \cite{note_Fokker_Planck}.
This in turn implies $\widetilde{V_\alpha V_\beta}\equiv\iint{dRdV V_\alpha V_\beta f_{eq}(R,V)}=k_BT/M\delta_{\alpha\beta}$, which, when used in Eq.(\ref{w_ia}) yields
\be
\widetilde{W_{ia}}P_i^{eq}=\widetilde{W_{ai}}P_a^{eq}\quad\text{for all }i,a\,. \label{detailed_balance_single}
\ee
Indeed, Eq.(\ref{w_ia}) implies
\be
\widetilde{W_{ia}}=\frac{2\pi k_BT}{M} d_{ia}\cdot d_{ia}^*e^{-\frac{\epsilon_{ia}}{2k_BT}}{\cal{L}}\l(\epsilon_{ia},\Gamma_{ia}\r)\,.
\ee
This is obtained using $\exp(x) \approx 1+x\approx$ for $|x|<1$ and the small magnitude of the exponent in Eq.(\ref{w_ia}).
Equation (\ref{detailed_balance_single}) is nothing but the detailed balance conditions, which says that the rates of the forward and backward non-adiabatic transitions between any pair of adiabatic electronic states, weighted by the probabilities of the initial and final states, are equal to each other.
With this relation, the r.h.s. of the master equations (\ref{dPidt}) vanishes and the quantum Boltzmann distribution $P_i^{eq} $ is stationary.

{\it (ii) Conservation properties.}
The Langevin equation implies the conservation over time of the number of classical particles and of the average momentum.
Similarly, the master equation implies the conservation of the normalization $\sum_i{P_i}$ over time.
As proved in Sec.~\ref{ProofEnergyConservation} , the scheme conserves the total energy $E(t)=\frac{M}{2}\sum_{\alpha=1}^{3N}{V_\alpha(t)^2}+\sum_i{P_i(t)\epsilon_i(t)}$ is conserved in the sense $\frac{d}{dt}\ll\! E(t)\!\gg=0$.

{\it (iii) Relation to other schemes.}
The scheme reduces to the Born-Oppenheimer approximation when all terms related to non-adiabatic couplings are dropped, which amounts to setting $\gamma$, $\xi$ and $\{W_{ia}\}$ to zero in Eq.(\ref{Langevinequation}) and (\ref{dPidt}).
When in our mathematical derivation the atomic degree of freedom are treated classically, as is the case, e.g., in the Ehrenfest method, the transition rates (\ref{w_ia}) become $\widetilde{W_{ia}}=2\pi\frac{k_BT}{M} d_{ia}\cdot d_{ia}^*{\cal{L}}\l(\epsilon_{ia},\Gamma_{ia}\r)$ (i.e. the exponential term disappears).
In this case, $W_{ia}=W_{ai}$ is satisfied, which violates the expected detailed balance relations (\ref{detailed_balance_single}).
Finally, we shall see that the new scheme can advantageously be cast in the form of the widely used fewest switches surface-hopping method proposed by Tully \cite{Tully1990}, but with switching probabilities that are not specified {\it ad-hoc}, but are instead derived.

\subsection{An algorithm}

Different algorithms can be envisioned to integrate the set of coupled equations (\ref{Langevinequation}) and (\ref{dPidt}).
Here, we find interesting to introduce an algorithm that is closely related to the popular ``fewest switches'' surface hopping method of Tully \cite{Tully1990}.
Below we assume that we have a practical way of generating all of the matrix elements $d_{ij}$.
For convenience, we closely follow Tully's original presentation (see steps 1 through 4 on page 1065 of \cite{Tully1990}) and adapt it to our purpose.
The algorithm propagates an ensemble of trajectories $(R(t),V(t),P_i(t))$.
Each trajectory moves on a weighted average potential energy surface, weighted by the occupation probabilities $P_i$, interrupted only by the possibility of sudden stochastic switches between electronic states.
Practically, the dynamics proceeds as follows:

Step 1.
Initial conditions $\{R_\alpha(0),V_\alpha(0),P_i(0)\}$ at time $t=0$ are assigned consistent with the physical conditions to be simulated (e.g., a thermal ensemble).

Step 2.
The classical equation of motion $M\ddot{R}_{\alpha}=F_{\alpha}^{BO}$ are integrated over a small time interval $\delta t$.

Step 3. The population $P_i$ of each state $i$ is then updated as follows.
A uniform random number $\xi_i$, $0<\xi_i<1$, is selected to determine whether a switch to any state $j$ will be invoked.
A switch to state $a$ occurs if $W_{ij}\delta t<\xi_i<W_{i,(j+1)}\delta t$.
In the notations of Tully (see Eq.(19) in \cite{Tully1990}), the switching probabilities in accordance to the new scheme are $g_{ij}=W_{ij}\delta t$, which differ from the original {\it ad-hoc} prescription. Note that the effective width of the ''delta''-function ${\cal L}(\epsilon_{ia}, \Gamma_{ia})$ in the expression for the rate in Eq. (\ref{w_ia}) is calculated ``on the fly'' according to Eq. (\ref{Gammafly}).

Step 4.
If a switch between a state $i$ and a state $j$ has occurred, an adjustment $\Delta V$ of atomic velocities $V$ must be made as follows in order to conserve energy, i.e.
\be
\frac{M V^2}{2}+\epsilon_i=\frac{M (V+\Delta V^{i})^2}{2}+\epsilon_j\,. \label{energy_conservation_constraint}
\ee
 the distribution of states into account, the velocity $V$ is adjusted to $V+\sum_i{P_i \Delta V^{i}}$.
As in \cite{Tully1990}, the adjustment $\Delta V$ is made in the direction of the non-adiabatic coupling $d_{ij}$ (here, we assume that we work with real-valued eigenstates $|i\rangle$, as can always be done in the absence of magnetic fields).
After the velocity adjustments have been made (if needed), return to step 2.
We observe that an advantage of this algorithm is that it does not necessitate the direct calculation of the coefficients $\gamma_{\alpha,\beta}$ and $B_{\alpha,\beta}$, but only requires the calculation of the non-adiabatic couplings $d_{ij}$ needed to evaluate the switching probabilities $W_{ia}$ and the adjustments $\Delta V^{i}$.
We refer the reader to the extensive literature on the calculation of non-adiabatic couplings $d_{ij}$; e.g., for widely used independent particle formulations such as density functional theory, see \cite{Maurer2016, Curchod2013} and references therein.

That this algorithm achieves a numerical solution of the scheme (\ref{Langevinequation})-(\ref{dPidt}) can be seen as follows.
Let $K(t)\equiv M \sum_i{P_i(t) \Delta V^{i}(t)}/\delta t$ denote the force change on the atoms described in Step 4.
The force $K(t)$ is a stochastic quantity, which results from the random switches between electronic states governed by the random numbers $\xi_i$.
Let $\langle\dots\rangle_\xi$ denote the average with respect to the uniform random numbers $\xi_i$ used in Step 3.
The energy conservation constraint (\ref{energy_conservation_constraint}) gives
\ben
\Delta V^{i}&=&-\frac{\epsilon_{ij}}{M d_{ij}\cdot V}\left[1+\frac{\epsilon_{ij}}{2M}\frac{d_{ij}\cdot d_{ij}}{|d_{ij}\cdot V|^2}\right]d_{ij}+O(\epsilon_{ij}^3)\,\,.
\een
With this expression, we find by straightforward algebra that $K(t)$ has the following statistical properties: with $\delta K_{\alpha}=K_\alpha-\langle K_\alpha\rangle_\xi$,
\bes
\be
\langle  K_\alpha(t)\rangle_\xi&=&M\sum_{\beta=1}^{3N}{\gamma_{\alpha,\beta}V_\beta(t)}\\
\left\langle \delta K_\alpha(t)\right\rangle_\xi&=&0\\
\left\langle \delta K_\alpha(t)\,\delta K_\alpha(t')\right\rangle_\xi&=&B_{\alpha,\beta}\delta(t-t')
\ee
\label{equations_K}
\ees
These properties hold provided $\delta t$ is small enough so that $W_{ij}\delta t\ll\! 1$, which is consistent with the primary motivation of Tully's fewest switches method that the electronic populations change with the minium number of hops \cite{Tully1990}.
The relations (\ref{equations_K}) justify the proposed algorithm as the sudden stochastic switches and associated velocity ``kicks'' reproduce, on average over the trajectories, the effect of the friction and random forces in the Langevin equation (\ref{Langevinequation}); indeed, $\langle  K_\alpha(t)\rangle_\xi$ equals the friction force in Eq.(\ref{Langevinequation}), while $\delta K_{\alpha}(t)$ has the same statistical properties as the stochastic force $\xi_\alpha(t)$ given in Eq.(\ref{xi_alpha}).

\section{Mathematical derivation and proofs} \label{Section_3}

\subsection{Effective Hamiltonian}

As remarked in the introduction, in order for a mixed quantum-classical scheme to satisfy the principle of detailed balance, it is essential to ensure that the canonical commutation relations of atomic variables be respected \cite{BlumBook,Abragam1961}.
This is not the case in the Ehrenfest method for the atomic positions are treated classically in the equations of electrons.
Instead, our derivation treats the atomic variables quantum mechanically before proceeding to the reductions leading to equations of evolution of the averaged atomic positions (\ref{Langevinequation}) and electronic populations (\ref{dPidt}).
We assume that the {\it total} wave function of the system can be written
\be
\Psi(r,R,t)=\phi\left(r,R,t;R_0(t)\right)\,, \label{Psi}
\ee
where the slow variations are carried by the parametric function $R_0(t)$ that is self-consistently set to equal the average atomic trajectory
\be
R_0(t)=\iint{R\,|\Psi(r,R,t)|^2drdR}\,. \label{definition_R0}
\ee
The ansatz (\ref{Psi}) recognizes that, as a consequence of the large atom to electron mass ratio and of the quasi-continuum of the density of electronic states, one can identify two well-separated time scales in the system: the slow, adiabatic time scale $T_\gamma$ of the classical atomic motion described by Eq.(\ref{definition_R0}), and the fast time scale $\tau_c\ll T_\gamma$ that characterizes the fluctuations of the interactions between electrons and atoms.
The ansatz (\ref{Psi}) nevertheless retains the quantum character of atomic variables $R$.
However, in order to account for the nearly classical character of ions, we assume that the dependence on $R$ of $\phi$ is strongly peaked around the averaged position $R_0(t)$, i.e.
\be
\langle \Psi(t)| (\hat{R}-R_0(t))^2 |\Psi(t)\rangle\ll\! R_0(t)^2 \,. \label{Taylorexpansion}
\ee
An illustrative example of a wave-function like Eq.(\ref{Psi}) is the product of an electronic wave-function times a Gaussian wave packet narrowly centered around $R_0(t)$ and of average momentum $M\dot{R}_0(t)$.
Extension to a statistical ensemble of states $\hat{\rho}$, e.g., a canonical ensemble, is straightforward and will be considered below.

The equations of motion for the atomic positions $R_0$ and electronic states populations outlined in Sec.~\ref{Section_2} are obtained by considering the propagation in time of the state $\Psi$ on an intermediate time scale $\tau_c\ll t\ll T_{\gamma}$ over which the electronic coherences vary (oscillate) widely, while the adiabatic atomic positions do not move appreciably.
Substituting Eq.(\ref{Psi}) into the Schr{\"o}dinger equation $i\hbar\frac{\partial}{\partial t}\Psi=\hat{H}\Psi$, we obtain the evolution equation of $\phi$ over the fast time scale,
\be
i\hbar \frac{\partial \phi}{\partial t}=\frac{\hat{P}^2}{2M}\phi+\hat{H}_e(\hat{r},\hat{R})\phi-i\hbar \dot{R}_0(t)\cdot\frac{\partial\phi}{\partial R_0(t)}\,. \label{dphidt}
\ee
The assumption (\ref{Taylorexpansion}) of spatially localized atomic positions  atomic positions allows one to approximate $\hat{H}\phi$ by the Taylor expansion,
\be
\hat{H}_e(\hat{r},\hat{R})\phi&\approx&\hat{H}_e(\hat{r},R_0(t))\phi \label{expanded_Hamiltonian}\\
&+&\left(\hat{R}-R_0(t)\right)\cdot\frac{\partial H_e}{\partial R}(\hat{r},R_0(t))\phi\,, \nn
\ee
so that Eq.(\ref{dphidt}) can be rewritten as the effective Schr{\"o}dinger equation
\be
i\hbar \frac{\partial}{\partial t}\phi=\hat{H}_{\rm eff}\phi\,, \label{dphidteff}
\ee
with the effective Hamiltonian \cite{Effecitve_Hamiltonian_note}
\be
\hat{H}_{\rm eff}&=&\frac{\hat{P}^2}{2M}+\sum_{n}{\epsilon_n|n\rangle\langle n|}-i\hbar\sum_{n,m}{\dot{R}_0(t)\cdot d_{nm}|n\rangle\langle m|}\nn\\
&&\hspace*{1.cm}+\sum_{n,m}{f_{nm}\cdot(\hat{R}-R_0(t))|n\rangle\langle m|}\,.\label{Heff}
\ee
Here we conveniently expressed the effective Hamiltonian in the orthonormal basis of adiabatic wave functions $|n[R_0(t)]\rangle$ defined above by Eq.(\ref{adiabatic_basis}) for $R(t)=R_0(t)$.
Below we shall often omit writing explicitly the dependence on $R_0(t)$ of the adiabatic basis and related quantities. Our scheme is obtained by considering the dynamics of $R_0$ and of the electronic state populations that result from the effective hamiltonian (\ref{Heff}).


\subsection{Initial conditions, statistical averaging}

It will be convenient to describe the electron dynamics in term of the electron density matrix $\hat{\rho}_e(r;R_0(t))$,
\ben
\langle r|\hat{\rho}_e(t;R_0(t))|r'\rangle=\int{\!dR\,\phi(r,R,t;R_0(t))\phi^*(r',R,t;R_0(t))}\,,
\een
and to expand the latter in terms of the adiabatic basis functions
\be
\hat{\rho}_e(t;R_0(t))&=&\sum_{i,j}{c_{ji}(t)|i[R_0(t)]\rangle\langle j[R_0(t)]|}\,. \label{rho_e_expansion}
\ee
For convenience, we introduce the notation (dropping dependencies),
\ben
\hat{c}_{ij}=|i\rangle\langle j|\,,
\een
so that $\hat{\rho}_e=\sum_{i,j}{c_{ji}\hat{c}_{ij}}$.
The coefficient $c_{ij}$ in the expansion (\ref{rho_e_expansion}) are given by
\be
c_{ij}=\langle j|\hat{\rho}_e|i\rangle={\rm Tr}\left[\hat{\rho}_e\hat{c}_{ij}\right]=\langle\Psi|\hat{c}_{ij}|\Psi\rangle\,. \label{cij_expressions}
\ee
The diagonal elements $c_{ii}$ are the electronic state populations, and the off-diagonal elements $c_{ij}$ define the coherences.
Our scheme outlined in Sec.~\ref{Section_2} describes the dynamics of the complete system in terms of the evolution of the averaged position $R_0(t)$ and of the populations $c_{ii}$.
The temporal evolution of initial coherences $c_{ij}$ to the next time step is not treated explicitly.
Indeed, as we shall see below, coherences fluctuate rapidly on a time scale smaller than the adiabatic time.
In the presence of a quasi-continuum of states, it is legitimate to neglect their influence on the evolution of the quantum population: this is the basis of the so-called secular approximation.
The remaining effect of coherences is on the atomic motions in the form of a rapidly fluctuating force that depends on the initial values of coherences $\rho_{ij}(t)$ only.
Here our lack of knowledge of the initial coherences is treated statistically: we assume that they are of the form $c_{ij}(0)e^{i(\phi_i-\phi_j)}$ where the phase factor $\phi_i$ is uniformly distributed in the interval $[0,2\pi]$.
If $\ll\!\dots\!\gg$ denotes the average of all phases $\{\phi_i\}$,
\ben
\ll\! c_{ij}(0)\!\gg&=&\ll\! c_{ii}(0)\!\gg \delta_{ij}\\
\ll\! c_{ij}(0)c_{kl}(0)\!\gg&=&\ll\! c_{ii}(0)\!\gg \delta_{i,l}\delta_{j,k}\,.
\een

\subsection{Evolution of atomic positions}

On the fast time scale, the effective Hamiltonian (\ref{Heff}) yields
\be
\frac{d\hat{R}}{dt}=\frac{\hat{P}}{M}\quad,\quad M\frac{d^2\hat{R}}{dt^2}=\frac{d\hat{P}}{dt}=-\sum_{i,j}{f_{ij}\hat{c}_{ij}} \label{dhatPdt}
\ee
Tracing over the total quantum state, Eq.(\ref{dhatPdt}) gives the equation of motion of $R_0$,
\be
M\frac{d^2R_0}{dt^2}=-\sum_{i,j}{f_{ij}c_{ij}(t)}\,.\label{d2R0dt2_intermediate}
\ee
The forces driving $R_0(t)$ are found by determining the temporal evolution of the $c_{ij}$'s.
The latter are obtained by integrating the evolution equations for the $\hat{c}_{ij}$ over the fast time scale driven by the effective Hamiltonian (\ref{Heff}),
\be
i\hbar\frac{d\hat{c}_{ij}}{dt}&=&\left[\hat{c}_{ij},\hat{H}_{eff}\right]\nn\\
&=&-\epsilon_{ij}\hat{c}_{ij}-i\hbar \sum_n{\left(\dot{R}_0\cdot d_{jn}\hat{c}_{in}-\dot{R}_0\cdot d_{ni}\hat{c}_{nj}\right)}\nn\\
&+&\sum_n{\left(f_{jn}\cdot\hat{X}_{in}-f_{ni}\cdot\hat{X}_{nj}\right)} \label{dhatcijdt}
\ee
where we defined $\hat{X}_{ij}=(\hat{R}-R_0(t))|i\rangle\langle j|$.
Averaging Eq.(\ref{dhatcijdt}) as in Eq.(\ref{cij_expressions}) to obtain the equation satisfied by $c_{ij}$ and integrating over time we find,
\be
\lefteqn{c_{ij}(t)=e^{i\epsilon_{ij}t/\hbar}c_{ij}(0)}&&\label{cij_for_atoms}\\
&-&\dot{R}_0(t)\cdot\int_0^{t}{dt'e^{i\epsilon_{ij}(t-t')/\hbar}\sum_n{\left(d_{jn}c_{in}(t')-d_{ni}c_{nj}(t')\right)}}\,.\nn
\ee
In deriving Eq.(\ref{cij_for_atoms}), we have neglected the last term in Eq.(\ref{dhatcijdt}) since its contribution to the non-adiabatic forces on the classical atoms is completely negligible.
As we will see in the next subsection, this is justified because the relevant states $n$ in Eq.(\ref{dhatcijdt}) satisfy the condition $3Nk_BT\gg\epsilon_{in},\epsilon_{jn}$.
Substituting the solution (\ref{cij_for_atoms}) in Eq.(\ref{d2R0dt2_intermediate}), the force driving the motion of $R_0$ is conveniently split into three parts such as
\be
M\frac{d^2R_0}{dt^2}&=&F^{BO}(t)+F^{\rm friction}(t)+\xi(t)\,. \label{generalized_Langevin_derived}
\ee
The first term
\ben
F^{BO}(t)=-\sum_{i}{c_{ii}(t)f_{ii}}=-\sum_{i}{c_{ii}\frac{\partial \epsilon_i}{\partial R_0}}
\een
is the traditional Born-Oppenheimer force.
The remaining two terms originate from the off-diagonal elements $c_{ij}(t)$, $i\neq j$ and are given by
\be
\lefteqn{F^{\rm friction}(t)=}&& \label{F_friction}\\
&&\sum_{i,j\neq i}{f_{ij}}\dot{R}_0\cdot \int_0^t{dt'e^{\epsilon_{ij}(t-t')/\hbar}\sum_n{\left(d_{jn}c_{in}(t')-d_{ni}c_{nj}(t')\right)}}\nn
\ee
and
\ben
\xi(t)=-\sum_{i,j\neq i}{f_{ij}e^{i\epsilon_{ij}t/\hbar}c_{ii}(0)}\,.
\een
They correspond, respectively, to the friction force and noise term of Eq.(\ref{Langevinequation}) that we discussed in Sec.~\ref{Section_2}.
Indeed,
\be
\lefteqn{F^{\rm friction}(t)}&&\nn\\
&\simeq&\sum_{i,j\neq i}{f_{ij}}\dot{R}_0\cdot d_{ji} \int_0^t{dt'e^{\epsilon_{ij}(t-t')/\hbar}\left(c_{ii}(t')-c_{jj}(t')\right)}\nn\\
\label{friction_step_1}\\
&\simeq&\sum_{i,j\neq i}{f_{ij}\left(c_{ii}(t)-c_{jj}(t)\right)\dot{R}_0\cdot d_{ji}\int_0^t{dt'e^{\epsilon_{ij}(t-t')/\hbar}}}\nn\\
\label{friction_step_2}\\
&\simeq&\pi\sum_{i,j\neq i}{f_{ij}\left(c_{ii}(t)-c_{jj}(t)\right)\dot{R}_0\cdot d_{ji} \delta\left(\epsilon_{ij}/\hbar\right)}\label{friction_step_3}\\
&=&-M\,\tensor{\gamma}\cdot\dot{R}_0(t) \label{F_friction_derived}
\ee
where we defined the matrix $\tensor{\gamma}$ of friction coefficients
\ben
\gamma_{\alpha\beta}=-\frac{\pi\hbar}{M}\sum_{i,j\neq i}{\frac{f_{ij}^\alpha\,f_{ji}^\beta}{\epsilon_{ij}}(c_{ii}-c_{jj})\delta(\epsilon_{ij})}\,,
\een
with $\alpha,\beta=1,\dots,3N$.
In deriving Eq.(\ref{F_friction_derived}), we have performed the following standard steps.
Firstly, in going from Eq.(\ref{F_friction}) to Eq.(\ref{friction_step_1}), we have used the secular approximation.
The latter consists in neglecting the off-diagonal term $i\neq n $ (and $j\neq n$) in Eq.(\ref{friction_step_1}) that, unlike the diagonal terms, oscillate rapidly at frequency $\epsilon_{in}/\hbar$ (see Eq.(\ref{cij_for_atoms})) and their overall contribution nearly cancels out as they interfere destructively for large enough finite times $t$.
The cancellation is most effective for denser density of states.
Secondly, in going from Eq.(\ref{friction_step_1}) to Eq.(\ref{friction_step_2}), we have replaced $c_{ii}(t')$ and $c_{jj}(t')$ by the values $c_{ii}(t)$ and $c_{jj}(t)$ at time $t$.
Indeed, in the contribution $\int_0^t{dt' \left[\sum_{j\neq i}{f_{ij}(\dot{R}_0\cdot d_{ji})e^{\epsilon_{ij}(t-t')/\hbar}}\right]c_{ii}(t')}$ (and similarly for the term involving $c_{jj}(t')$), the only values of $c_{ii}(t')$ to contribute significantly to the integral are those which correspond to $t'$ very close to $t$ since the sum in the square bracket practically interfere destructively for a quasi-continuum of states at soon as $t-t'\gg \hbar/\Delta$, where $\Delta$ is the energy ``width'' of $F(\epsilon_i)=\sum_{j\neq i}{f_{ij}(\dot{R}_0\cdot d_{ji})e^{\epsilon_{ij}(t-t')/\hbar}}$, i.e. the order of magnitude of the $\epsilon_i$ variation needed for the sum to change $F(\epsilon_i)$ significantly.
Lastly, in going from Eq.(\ref{friction_step_2}) to Eq.(\ref{friction_step_3}), we have used
\be
\int_0^t{dt' e^{i\epsilon_{in}(t-t')/\hbar}}&=&e^{i\epsilon_{in}t/2}\frac{\sin\left(\epsilon_{in}t/2\right)}{\epsilon_{in}/2}\nn\\
&\approx&\pi\hbar\delta(\epsilon_{in})+i\hbar{\cal{P}}\frac{1}{\epsilon_{in}}\,. \label{approximate_delta}
\ee
at large enough $t$.
Note that is is not necessary to let $t$ approach infinity in order to use (\ref{approximate_delta}) in Eq.(\ref{friction_step_2}) but it suffices for $\hbar/t$ to be smaller than the energy ``width'' of $F(\epsilon_i)$ discussed above.
As for the term $\xi(t)$, it corresponds to a delta-correlated Gaussian random force characterized by the relations Eqs.(\ref{xi_alpha}).
Indeed, the properties (\ref{xi_alpha}) imply
\be
\lefteqn{\ll\! \xi_\alpha(t)\xi_\beta(t')\!\gg}&&\nn\\
&=&\sum_{a,b}{}^{'}{\sum_{k,l}{}^{'}{f_{ab}^{\alpha}f_{kl}^{\beta}e^{i(\epsilon_{ab}t+\epsilon_{kl}t')/\hbar}\ll\! c_{ab}(0)c_{kl}(0)\!\gg}}\nn\\
&=&\sum_{a,b}{}^{'}{f_{ab}^{\alpha}f_{ba}^{\beta}e^{i\epsilon_{ab}(t-t')/\hbar}c_{aa}(0)}\nn\\
&\approx& B_{\alpha,\beta}\delta(t-t')\label{noise_property}
\ee
with
\ben
B_{\alpha,\beta}&=&\int{d(t-t')\sum_{a,b}{}^{'}{f_{ab}^{\alpha}f_{ba}^{\beta}e^{i\epsilon_{ab}(t-t')/\hbar}c_{aa}(0)}}\\
&=&\pi\sum_{i\neq j}{(P_i+P_j)f_{ij}^{\alpha}f_{ji}^{\beta}\delta(\epsilon_{ij}/\hbar)}
\een
More generally, Eq.(\ref{noise_property}) can be written $\ll\!\xi_{\alpha}(t) \xi_{\beta}(t') \!\gg=B_{\alpha,\beta}g(t-t')$, where $g(t)$ is an even, normalized function of width $\tau_c$.
If, as assumed here, the atomic motions are integrated over time steps $\delta t>\tau_c$, $g$ can be approximated by a delta function as in Eq.(\ref{noise_property}).

\subsection{Evolution of electronic populations}

We consider again the evolution equations (\ref{dhatcijdt}) of the operators $\hat{c}_{ij}$ over the fast time scale governed by the effective Hamiltonian. 
In order to get the rate equations for the electronic populations $\langle\hat{c}_{ii}\rangle$ we iterate Eqs.(\ref{dhatcijdt}). That is, we formally solve Eq.(\ref{dhatcijdt}) for operators ${\hat c}_{ij}(t)$,
\be
{\hat c}_{ij}(t)&=& {\hat c}_{ij}(0)\,e^{-i\epsilon_{ij}t/\hbar}\label{dcijdt}\\
&+&\int_0^t dt'\, e^{i\epsilon_{ij}(t-t')/\hbar}\sum_{n}\left[d_{jn}{\hat V}_{in}(t')-d_{ni}{\hat V}^\dag_{jn}(t')\right]\,,\nn
\ee
where
\be\nonumber
{\hat V}_{in}= i {\dot R_0} {\hat c}_{in} + (1/\hbar)\epsilon_{ij}{\hat c}_{in}\,\delta {\hat R}\,,
\ee
$\delta {\hat R} = {\hat R}-R_0(t)$, and the ``dagger'' stands for hermitian conjugation, and substitute these expressions back into Eqs. (\ref{dhatcijdt}). In doing so we should keep in mind that the order of operators $\delta {\hat R}$ and $\hat{c}_{in}$ matters. While these operators commute at the same time, they do not commute when taken at different $t$'s due to the non-commutativity between $\delta {\hat R}$ and ${\hat P}$. Then, applying the same approximations as we did in deriving equations of motion for the ions in the previous section, i.e., dropping the rapidly oscillating terms  as well as the terms linear in $\delta {\hat R}$ terms (which average to zero), we arrive at the following equation of motion for the electronic populations $c_{ii}=\langle {\hat c}_{ii}\rangle$,
\be
\frac{d c_{ii}}{dt}=\sum_{n\ne i}{\left(W_{in}c_{nn}-W_{ni}c_{ii}\right)} + \eta_i(t) \label{rate_equation_derived}
\ee
with the rates
\be
W_{in}&=& d_{in}^\alpha d_{in}^\beta\int_0^t dt'\Big\{2{\rm Re}[{\dot R}_{0\alpha}(t){\dot R}_{0\beta}(t')\,e^{i\epsilon_{in}(t-t')/\hbar}]\nonumber\\
&+&(\epsilon_{in})^2\Big[\langle \delta {\hat R}_\alpha(t)\delta {\hat R}_\beta(t')\rangle e^{i\epsilon_{in}(t-t')/\hbar}\nonumber\\
&+&\langle \delta {\hat R}_\beta(t')\delta {\hat R}_\alpha(t)\rangle e^{i\epsilon_{in}(t'-t)/\hbar}\Big]\Big\}\,,\label{rate1}
\ee
and the noise terms
\be
\eta_i(t) &=& \sum_n\Big\{ d_{in}\cdot \left[ i{\dot R_0} + (1/\hbar)\epsilon_{in}\,\delta {\hat R} \right]{\hat c}_{in}(0)\,e^{-i\epsilon_{in}t/\hbar}\nonumber\\
&-& d_{ni}\cdot \left[ i{\dot R_0} + (1/\hbar)\epsilon_{ni}\,\delta {\hat R} \right]{\hat c}_{in}(0)\,e^{i\epsilon_{in}t/\hbar}\Big\}\,.\label{noise}
\ee
The first contribution to the rates in the rhs of Eq. (\ref{rate1}) is the dominating term associated with the classical motion of atoms. The second term is smaller, but plays an important role. To see this, let's integrate this second term by parts (over $t'$). The boundary terms vanish: The $t'=t$ contribution exactly cancels, while $t'=0$ contributions, $\langle \delta {\hat R}_\alpha(t)\delta {\hat R}_\beta(0)\rangle e^{i\epsilon_{in}t/\hbar}$ and $\langle \delta {\hat R}_\alpha(0)\delta {\hat R}_\beta(t)\rangle e^{-i\epsilon_{in}t/\hbar}$ vanish in the limit of sufficiently large $t$. The remaining term is
\be
i(d_{in}^\alpha d_{in}^\beta \epsilon_{in}/M) \int_0^t dt' \Big[\langle \delta {\hat R}_\alpha(t)\delta {\hat P}_\beta(t')\rangle e^{i\epsilon_{in}(t-t')/\hbar}\nonumber\\
-\langle \delta {\hat P}_\beta(t')\delta {\hat R}_\alpha(t)\rangle e^{i\epsilon_{in}(t'-t)/\hbar} \Big]\,,
\ee
where we have used that $d\delta{\hat R}/dt = \delta{\hat P}/M$. Furthermore, since this term is small compared to the first, ``classical'' term in Eq. (\ref{rate1}), we can set $t=t'$ in the correlation functions $\langle \delta {\hat R}_\alpha(t)\delta {\hat P}_\beta(t')\rangle$, etc. Then, the virtue of coordinate-momentum commutation relation $[\hat P_\alpha,\,\hat R_\beta]=i\hbar\,\delta_{\alpha\beta}$, the above expression is written as
\be
(d_{in}^\alpha d_{in}^\alpha \epsilon_{in}/M)\int_0^t dt'\Big\{\hbar\,{\rm Re}[e^{i\epsilon_{in}(t-t')/\hbar}]\nonumber\\
+i\langle\{\delta {\hat R}_\alpha(t),\delta {\hat P}_\beta(t)\}\rangle\, {\rm Im}[ e^{i\epsilon_{in}(t-t')/\hbar}]\Big\}\,.\label{asymrate}
\ee

The anti-commutator $\delta {\hat R}_\alpha,\delta {\hat P}_\beta\}$, when averaged relative to a spatially localized atomic state is small.  For instance, for normalized Gaussian wave-functions $\Psi(R)=e^{-(R-R_0)^2/2\sigma^2}e^{iP_0\cdot(R-R_0)/\hbar}/\pi^{1/4}\sqrt{\sigma}$ (written here in one-dimension for simplicity), it is equal to zero. Also, its contribution, unlike that for the first term in Eq. (\ref{asymrate}), is imaginary and so it can only contribute to the renormalization of energy differences, e.g. $\epsilon_{in}$, but not to the rates. Thus we get
\be
W_{in}&=& 2\,d_{in}^\alpha d_{in}^\beta {\rm Re}\Big\{\int_0^t dt'\Big[{\dot R}_{0\alpha}(t){\dot R}_{0\beta}(t')\nonumber\\
&+&\delta_{\alpha\beta}{\epsilon_{in}\over 2M} \Big] e^{i\epsilon_{in}(t'-t)/\hbar}\Big\}\,.\label{rate2}
\ee

The ${\dot R}_{0\alpha}(t){\dot R}_{0\beta}(t')$  term in the rhs of Eq. (\ref{rate2}) can be transformed to a
\be\label{velocity}
{\dot R}_{0\alpha}(t){\dot R}_{0\beta}(t) - (t-t')B_{\alpha,\beta}/M^2\,,
\ee
where $B_{\alpha,\beta}$ is given by Eq. (\ref{noise_property}). Indeed, by writing
\be\nonumber
{\dot R}_{0\alpha}(t') = {\dot R}_{0\alpha}(t) + \int_t^{t'}{\ddot R}_{0\alpha}(t_1)\,dt_1   \,,
\ee
and taking into account that on the short time scale the classical atomic coordinate obeys $M{\ddot R}_{0\alpha}=\xi_\alpha(t)$, we can average $(1/M)\int_t^{t'}{\dot R}_{0\alpha}(t)\xi_\beta(t_1)\,dt_1$  over the white noise, obtaining the second term in Eq. (\ref{velocity}). Furthermore, exponentiating $(t'-t)B_{\alpha,\beta}/M^2 + \epsilon_{in}/(2M)$, we obtain that
\be
W_{in}&=& 2 [d_{in}\cdot {\dot R}_0(t)]^2\,e^{{\epsilon_{in}d_{in}^2\over 2M[d_{in}\cdot {\dot R}_0(t)]^2}}\nonumber\\
&\times&{\rm Re} \Big\{\int_0^t dt'\,e^{(i\epsilon_{in}/\hbar -\Gamma_{in})(t'-t)}\Big\}\,,\label{rate3}
\ee
where
\be\label{gamma1}
\Gamma_{in}= d_{in}^\alpha d_{in}^\beta B_{\alpha,\beta}(t)/[M d_{in}\cdot {\dot R}_0(t)]^2\,.
\ee
Thus we can write the rate as
\be\label{rate4}
W_{in}= 2\pi [d_{in}\cdot {\dot R}_0(t)]^2\,e^{{\epsilon_{in}d_{in}^2\over 2M[d_{in}\cdot {\dot R}_0(t)]^2}}\,{\cal{L}}\l(\epsilon_{in},\Gamma_{in}\r)\,,
\ee
with the "delta function" having finite width due to the noise induced by the electrons on the atomic motions.

The additive noise $\eta_i(t)$ in Eqs. (\ref{rate1}, \ref{noise}) has zero average over the ensemble of initial conditions ${\hat c}_{in}(0)$, and it is``delta''-correlated on atomic timescale. A straightforward calculation analogous to the derivation of expression (\ref{rate4}) leads to
\be\label{noiseelectron}
\langle\eta_i(t)\eta_j(t')\rangle = \Big[\sum_{n\neq i}(W_{in}+W_{ni})\Big]\, \delta_{ij}\,\delta(t-t')\,. 
\ee
Eqs. (\ref{rate_equation_derived}, \ref{noiseelectron}) imply that the electronic dynamics is Markovian, with the electronic subsystem spontaneously ``hopping'' between states with gradually varying energies $\epsilon_i$. Upon averaging the stochastic rate equations (\ref{rate1}) over the noise configurations one obtains the master equation (\ref{dPidt}) with $P_i=\langle c_{ii} \rangle\equiv\,\ll {\hat c}_{ii}\gg$.

\subsection{Proof of energy conservation property} \label{ProofEnergyConservation}

The total energy of the system at time $t$ in the state $\phi(r,R,t;R_0(t))$ along a trajectory $R_0(t)$ is
\ben
E_{tot}(t)=\frac{M\dot{R}_0^2}{2}+\sum_i{\epsilon_iP_i}\,.
\een
We will show that the ensemble averaged energy $\ll\! E_{tot}(t)\!\gg$ is conserved over time, i.e.
\ben
\frac{d\ll\! E_{tot}(t)\!\gg}{dt}=0\,. \label{dE_dt_derived}
\een

Firstly, taking the dot product of the equation of motion (\ref{generalized_Langevin_derived}) by the vector $\dot{R}_0(t)$ of atomic velocities, we obtain
\be
\lefteqn{\frac{M}{2}\frac{d}{dt}{\dot R}_0^2+\sum_i{\frac{d\epsilon_i}{dt}P_i}}&&\label{atomenergy}\\
&&=2\pi\hbar\sum_{i,j}{}^{'}{P_{i}\frac{|\dot{R}_0\cdot f_{ij}|^2}{\epsilon_{ij}}\delta(\epsilon_{ij})}+\dot{R}_0\cdot\xi\,.\nonumber \label{R0dottimesLangevin}
\ee
Secondly, multiplying the rate equation (\ref{rate_equation_derived}) by the electronic energy $\epsilon_i$, and summing over all states $i$, we find
\be
\lefteqn{\sum_{i}{\epsilon_i \frac{dP_i}{dt}}}&&\label{electronenergy}\\
&=&-2\pi\hbar\sum_{ij}{}^{'}{P_i\frac{|\dot{R}_0\cdot f_{ij}|^2}{\epsilon_{ij}}\delta(\epsilon_{ij}) }-\frac{1}{2M}\sum_{ij}{}^{'}{B_{ij}}\,.\nonumber
\ee
Summing the previous two equations yields
\ben
\frac{dE_{tot}}{dt}=-\frac{1}{2M}\sum_{ij}{}^{'}{B_{ij}}+\dot{R}\cdot\xi\,.\label{energytot}
\een
This equation averages out to zero upon averaging over the initial phases to yield the desired result.
Indeed, using the stochastic equation (\ref{generalized_Langevin_derived}) and the property (\ref{noise_property}), we find,
\ben
\l\la \dot{R}(t)\cdot\xi(t)\r\ra=\l\la \int_0^t{dt'\,\frac{\xi(t')}{M}}\cdot\xi(t) \r\ra=\frac{1}{2M}\sum_{ij}{}^{'}{B_{ij}}\,.
\een
for $\tau_c\ll t\ll T_\gamma$.

\subsection{Linewidth}

The equations for the rates derived in this section, e.g. Eq. (\ref{rate4}) contain a broadened $\delta$-function of the energy difference $\epsilon_{in}=\epsilon_i-\epsilon_n$ with the width given by Eq.(\ref{gamma1}). The latter equation contains the atomic noise correlator $B_{\alpha,\beta}$, which gradually varies with time due to the changing atomic positions in the course of the evolution. The instantaneous (on the timescale of atomic motions) value of $B_{\alpha,\beta}$ is given Eq. (\ref{fluctuation_dissipation}), which, in its turn, contains a sum over delta-functions. In our derivation in section (III. C) the values for the width of the delta functions in Eq. (\ref{fluctuation_dissipation}) was not specified for two reasons: Firstly, it can not be easily derived within the framework of the calculation presented in this paper. Presumably, it can be carried out by analyzing the dynamics of higher correlators, $\langle {\hat R} {\hat c}_{ij} \rangle$, $\langle {\hat P} {\hat c}_{ij} \rangle$, etc., which lies beyond the scope of this calculation. Secondly, since the delta functions in Eqs. (\ref{friction_gamma}, \ref{fluctuation_dissipation}) are effectively under the integrations over the energies, their precise widths should not matter (unless the sums in Eqs.(\ref{friction_gamma}, \ref{fluctuation_dissipation}) contain a finite number of terms).

The energy conservation property derived in the previous subsection, however, dictates that the widths of the``delta''-functions in Eqs. (\ref{friction_gamma}, \ref{fluctuation_dissipation}) and in Eq. (\ref{rate4}) should be the same (as was saliently implied in Eqs. (\ref{gamma1}, \ref{rate4}), where we have used the same $\Gamma_{in}$ as in Eqs. (\ref{friction_gamma}, \ref{fluctuation_dissipation}, \ref{call})). Indeed, had $\Gamma_{in}$'s in these expressions were different, the first terms in the rhs of Eq. (\ref{atomenergy}) and Eq. (\ref{electronenergy}) would not cancel each other in Eq. (\ref{energytot}) (again, for a finite number of electronic states). Therefore, Eq. (\ref{gamma1}) can be rewritten as a self-consistency equation for determining $\Gamma_{in}$, e.g. Eq. (\ref{gammaself1}).

The rhs of Eq. (\ref{gammaself1}) can be easily estimated if we assume that the number of atomic degrees of freedom is large. Then the expression $(M d_{in}\cdot V)^2$ in Eq. (\ref{gammaself1}) self-averages to give $M^2 d_{in}^2 \langle V_\alpha^2\rangle = M d_{in}^2 T_i$, where $T_i$ is the temperature of the atomic degrees of freedom. Furthermore, the sum over $j, j'$, i.e., $B_{\alpha,\beta}$,  by virtue of the fluctuation-dissipation theorem, can be estimated to give $\sim 2M\gamma T_e\delta_{\alpha\beta}$, where $\gamma$ is the typical atomic friction coefficient and $T_e$ is the electronic temperature. Thus we obtain that
\be
\Gamma \sim 2\,{T_e\over T_i}\,\gamma\,, \label{Gammaest}
\ee
and so in the equilibrium, i.e., for $T_e\sim T_i$, the linewidths $\Gamma_{in}$ are of the order of the atomic friction coefficient.           

In the surface hopping algorithm described in section II. C the values of $\Gamma_{in}$ should be calculated ``on the fly'', i.e., when the electronic subsystem is in a particular state $i_0$. In that case we should replace $P_j$ in Eq. (\ref{gammaself1}) by $\delta_{i_0j}$, which leads to the following self-consistency equation for $\Gamma_{i_0n}$,
\be
\Gamma_{i_0n}&=& {2\hbar^2\over (M d_{i_0n}\cdot V)^2}\sum_{j\neq i_0}
\frac{(d_{i_0n}\cdot f_{i_0j})^2\Gamma_{i_0j}}{(\epsilon_{i_0j})^2+(\hbar\Gamma_{i_0j})^2}\,.\label{Gammafly}
\ee

It should be pointed out that $\Gamma$'s evaluated according to Eqs. (\ref{gammaself1}, \ref{Gammafly}) are not the actual electronic line widths. The latter should include effects related to the disorder caused by the randomness in atomic positions and thus the fluctuations in the energies $\epsilon_i$ of the electronic states. Such disorder effects can be accounted for by running a {\it bundle} of trajectories with different initial conditions (rather than a single trajectory). Averaging the fluctuations in $\epsilon_i$'s over the ensemble of these trajectories will produce the linewidths of the order of $\tau_c^{-1}$, discussed in Sec. II. A. These linewidths define the rate of decay of electronic correlations or coherences as well as the transport time. On the contrary, the partial widths $\Gamma_{in}$ ($\Gamma\sim T_\gamma^{-1}$) are associated with the atomic jittering around a {\it single} trajectory and therefore define the linewidths for the transitions associated with this trajectory.

\section{Summary} \label{Section_4}

In summary, we have presented a scheme for carrying out non-adiabatic molecular dynamics simulations of systems where a quasi-continuum of electronic excitations of arbitrarily low energy is available to couple with the ionic motions.
The scheme, which is derived from first principles and does not rely on {\it ad-hoc} parameters, naturally satisfies the principle of detailed balance and, therefore, can properly describe non-equilibrium dynamics.
A numerical algorithm is proposed in the form of the widely used fewest switches surface-hopping algorithm but with switching probabilities that are not specified {\it ad-hoc} like in the original algorithm but are instead derived.
The present approach could greatly increase the number of processes amenable to realistic simulation by molecular dynamics in several research areas.

\section{Acknowledgments}
This work was performed under the auspices of the United States Department of Energy under Contract DE-AC52-06NA25396 and supported by LDRD Grant No. 20170490ER and 20170460ER.


\begin{references}
\bibitem{Tully1998} J.C. Tully, {\it Faraday Discuss.} {\bf 110}, 407 (1998); J.C. Tully, in {\it Modern Methods for Multidimensional Dynamics Computations in Chemistry}, ed. D.L. Thomson (World Scientific, 1998), p.34.
\bibitem{Tully2012} J.C. Tully, {\it J. Chem. Phys.} {\bf 137}, 22A301 (2012).
\bibitem{Tully1990} J.C. Tully, {\it J. Chem. Phys.} {\bf 93}, 1061 (1990).
\bibitem{Stella2007} L. Stella, M. Meister, A. J. Fisher, and A.P. Horsfield, {\it J. Chem. Phys.} {\bf 127}, 214104 (2007).
\bibitem{Parandekar2005} P.V. Parandekar and J.C. Tully, {\it J. Chem. Phys.} {\bf 122}, 094102 (2005).
\bibitem{Bastida2006} A. Bastida, C. Cruz, J. Zuniga, A. Requena, and B. Miguel, {\it Chem. Phys. Lett.} {\bf 417}, 53 (2006). In this reference, an ad-hoc modification of the Ehrenfest equations is used to enforce the correct equilibrium distribution of a quantum system coupled to a classical bath.
\bibitem{Wodtke2004} A. M. Wodtke, J. C. Tully, and D. J. Auerbach, {\it Int. Rev. Phys. Chem.} {\bf 23}, 513 (2004).
\bibitem{Askerka2016} M. Askerka, R.J. Maurer, V.S. Batista and J.C. Tully, {\it Phys. Rev. Lett.} {\bf 116}, 217601 (2016).
\bibitem{Li1995} Y. Li and G. Wahnstr{\"o}m, {\it Phys. Rev. B} {\bf 51}, 12233 (1995).
\bibitem{Race2010} C.P. Race, D.R. Mason, M.W. Finnis, W.M.C. Foulkes, A.P. Horsfield, and A.P. Sutton, {\it Rep. Prog. Phys.} {\bf 73}, 116501 (2010).
\bibitem{Correa2012} A.A. Correa, J. Kohanoff, E. Artacho, D. S{\'a}nchez-Portal, and A. Caro, {\it Phys. Rev. Lett.} {\bf 108}, 213201 (2012).
\bibitem{IPAM2012} {\em Frontiers and Challenges in Warm Dense Matter}, Series: Lecture Notes in Computational Science and Engineering, 96, edited by F. Graziani, M.P. Desjarlais, R. Redmer, and S.B. Trickey (Springer 2014)
\bibitem{Pradhan2016} E. Pradhan, R.J. Magyar, and A.V. Akimov, {\it Phys. Chem. Chem. Phys.} {\bf 18}, 32466 (2016).
\bibitem{Cho2016} B. I. Cho, T. Ogitsu, K. Engelhorn, A. A. Correa, Y. Ping, J. W. Lee, L. J. Bae, D. Prendergast, R. W. Falcone, and P. A. Heimann, {\it Scientific Reports} {\bf 6}, 18843 (2016).
\bibitem{Valero2012} R. Valero and D. G. Truhlar, {\it J. Chem. Phys}. {\bf 137}, 22A539 (2012).
\bibitem{BlumBook} K. Blum {\it Density Matrix Theory and Applications, 3rd Edition} (Springer-Verlag, 2012), p. 287-288.
\bibitem{Abragam1961} A. Abragam, {\it The Principles of Nuclear Magnetism} (Clarendon Press, Oxford, 1961).
\bibitem{Note_extension} The presentation can be extended to any fast quantum coordinates $r$ in interaction with slow classical coordinates $R$.
\bibitem{Curchod2013} B.F.E. Curchod, U. Rothlisberger, and I. Tabernelli, {\it ChemPhysChem} {\bf 14}, 1314 (2013).
\bibitem{Maurer2016} R.J. Maurer, M. Askerka, V.S. Batista, and J.C. Tully, {\it Phys. Rev. B} {\bf 94}, 115432 (2016).
\bibitem{note_on_delta_time} In general $\ll\!\xi_{\alpha}(t) \xi_{\beta}(t')\!\gg=B_{\alpha,\beta}g(t-t')$, where $g(t)$ is an even, normalized function of width $\tau_c$. If, as assumed here, the atomic motions are integrated over time steps $\delta t>\tau_c$, $g$ can be approximated by a delta function.
\bibitem{non_on_gamma} Alternatively, in practice, one could set all $\Gamma_{ij}$ to the same, empirically chosen value $\Gamma$, as is commonly done.
\bibitem{dAgliano1975} E.G. d'Agliano, P. Kumar, W. Schaich, and H. Suhl {\it Phys. Rev. B} {\bf 11}, 2122 (1975).
\bibitem{HeadGordonTully1995} M. Head-Gordon and J.C. Tully {\it J. Chem Phys.} {\bf 103}, 10137 (1995).
\bibitem{DaligaultMozyrsky2007} J. Daligault and D. Mozyrsky, {\it Phys. Rev. E} {\bf 75}, 026402 (2007).
\bibitem{Lu2012} J.-T. L\"u, M. Brandbyge, and P. Hedegard, T. Todorov, and D. Dundas {\it Phys. Rev. B} {\bf 85}, 245444 (2012).
\bibitem{Dou2017} W. Dou, G. Miao, and J. E. Subotnik, {\it Phys. Rev. Lett.} {\bf 119}, 046001 (2017).
\bibitem{note_Fokker_Planck} The swarm of trajectories generated by Eq.(\ref{Langevinequation}) can be described by a probability distribution function $f(R,V,t)$ that satisfies the Fokker-Planck equation
\ben
\lefteqn{\frac{\partial f}{\partial t}+V\cdot\frac{\partial f}{\partial R}+\frac{F^{BO}}{M}\cdot\frac{\partial f}{\partial V}}&&\\
&&=\sum_{\alpha,\beta}^{3N}{\frac{\partial}{\partial V_\alpha}\left[\gamma_{\alpha,\beta}\left(V_{\beta}f\right)+\frac{B_{\alpha,\beta}}{2M^2}\frac{\partial}{\partial V_{\beta}}f\right]}\,.
\een
$f_{eq}(R,V)=\exp\left[-\frac{1}{k_BT}\left(\frac{MV^2}{2}+\phi_{B0}(R)\right)\right]/{\cal{Z}}_{cl} $ is an equilibrium solution of this equation.
\bibitem{Effecitve_Hamiltonian_note}
In general,
\ben
\hat{H}_{\rm eff}&=&\frac{\hat{P}^2}{2M}+\sum_{n}{\left[\epsilon_n|n\rangle\langle n|-i\hbar|n\rangle\frac{\partial}{\partial R_0}\langle n|\right]}\\
&+&\sum_{n,m}{\left[f_{nm}\cdot(\hat{R}-R_0(t))-i\hbar\dot{R}_0(t)\cdot d_{nm}\right]|n\rangle\langle m|}
\een
but the term $-i\hbar|n\rangle\frac{\partial}{\partial R_0}\langle n|$ does not affect the next steps.
\end{references}
\end{document}